\documentclass[runningheads]{llncs}
\usepackage[misc]{ifsym}
\usepackage{epstopdf}
\usepackage{graphicx}
\usepackage{booktabs}
\usepackage{multirow}
\usepackage{bbding}
\usepackage{amssymb}
\usepackage{amsmath}
\begin{document}
\title{3D SA-UNet: 3D Spatial Attention UNet with 3D Atrous
Spatial Pyramid Pooling for White Matter Hyperintensities
Segmentation\thanks{$^{1}$https://github.com/clguo/3DSAUNet \\
$^{2}$https://github.com/hjkuijf/wmhchallenge/blob/master/evaluation.py }}

\titlerunning{3D SA-UNet: 3D Spatial Attention UNet}
%

\author{Changlu Guo $^{(\textrm{\Letter})}$ 
}

\authorrunning{Guo et al.}
\institute{
Independent Researcher (work conducted before current affiliation)\\
\email{clguo.ai@gmail.com}
}

\maketitle              
\begin{abstract}
White Matter Hyperintensity (WMH) is an imaging feature related to various diseases such as dementia and stroke. Accurately segmenting WMH using computer technology is crucial for early disease diagnosis. However, this task remains challenging due to the small lesions with low contrast and high discontinuity in the images, which contain limited contextual and  spatial information. To address this challenge, we propose a deep learning model called 3D Spatial Attention U-Net (3D SA-UNet) for automatic WMH segmentation using only Fluid Attenuation Inversion Recovery (FLAIR) scans. The 3D SA-UNet introduces a 3D Spatial Attention Module that highlights important lesion features, such as WMH, while suppressing unimportant regions. Additionally, to capture features at different scales, we extend the Atrous Spatial Pyramid Pooling (ASPP) module to a 3D version, enhancing the segmentation performance of the network. We evaluate our method on publicly available dataset and demonstrate the effectiveness of 3D Spatial Attention Module and 3D ASPP in WMH segmentation. Through experimental results, it has been demonstrated that our proposed 3D SA-UNet model achieves higher accuracy compared to other state-of-the-art 3D convolutional neural networks. The code can be found in $Github^{1}$.

\keywords{White Matter Hyperintensity Segmentation\and Deep Learning \and Spatial Attention \and 3D SA-UNet \and ASPP}
\end{abstract}

\section{Introduction}
In the field of medical imaging, Small Vessel Disease (SVD) is a pathological condition that affects the small blood vessels of the brain and is commonly seen in patients with conditions such as aging, stroke, and dementia \cite{Wardlaw}. One of the hallmarks of this disease is White Matter Hyperintensity (WMH), which appears as abnormally bright areas in Magnetic Resonance Imaging (MRI) images. The occurrence of WMH is typically associated with microvascular damage to brain tissue, reflecting potential neurodegenerative changes.

With the advancement of medical imaging technology, the quantification of WMH—encompassing its location, shape, and volume in MRI images—has become an important means of diagnosing and assessing neurodegenerative diseases. This kind of quantitative analysis can not only assist clinicians in making more accurate diagnoses but also provide vital information for prognosis evaluation and treatment monitoring. For instance, in stroke patients, the quantification of WMH can help predict the risk of recurrence and the course of recovery; in dementia patients, it can aid in distinguishing between different types of dementia, such as Alzheimer's disease and vascular dementia.

However, the assessment of WMH currently relies mainly on visual grading by trained radiologists, a method widely used in clinical practice but with clear limitations. Visual grading is not only time-consuming but also subject to subjective judgment, with potential differences in assessment between different physicians. Moreover, as the number of patients increases, the workload of physicians also increases, further exacerbating the issues of inconsistency and time delays in assessment.

To overcome these limitations, automated WMH segmentation techniques have emerged. Automated segmentation utilizes computer vision and machine learning algorithms to quickly and objectively identify and quantify WMH from MRI images. This method not only reduces human error and improves the consistency and accuracy of assessments but also significantly increases work efficiency, allowing physicians to devote more time and energy to the diagnosis and treatment of diseases. With the continuous advancement of advanced technologies such as deep learning, the accuracy and robustness of automated WMH segmentation are also continuously improving, demonstrating great potential in clinical applications.

With the widespread application of deep learning in medical image processing, the automatic segmentation of WMH has also benefited, with popular models like U-Net \cite{unet} being employed for this task. Guerreroa et al.\cite{Guerrero} proposed a network called uResNet, which combines the strengths of ResNet\cite{resnet} and U-Net for segmenting hyperintensities using 2D image patches. Li et al.\cite{Li} proposed an ensemble method that combines multiple U-Nets and achieved first place in the WMH Segmentation Challenge at MICCAI 2017\cite{Kuijf}.  Vaanathi et al. \cite{Sundaresan} devised an ensemble triplanar network that integrated the predictions from three distinct planes of MRI images to achieve a precise segmentation of white matter hyperintensities (WMH). However, it is easy to find that few 3D convolutional neural networks have been applied to this work, mainly due to the low contrast and high discontinuity of small lesion areas in WMH images, containing poor spatial and contextual information, as well as the poor imaging resolution and large spatial resolution variation of these images along the $z$-direction, which limit the use of 3D deep learning models\cite{Li}. To tackle the aforementioned challenges, we employ a bold design strategy by using a 3 $\times$ 3 $\times$ 1 convolution kernel instead of the traditional 3 $\times$ 3 $\times$ 3 convolution kernel in both the 3D encoder and decoder, allowing the network to focus on image information in the $x$ and $y$ directions. Furthermore, in the bottleneck block between the encoder and decoder, we retain the design of 3D convolution to preserve the network's ability to capture essential 3D features. To further enhance the network's capability to capture multi-scale information, we introduce an extended version of 3D ASPP within the bottleneck block. This extension enables the network to learn and integrate multi-scale information more effectively. Finally, unlike the skip connections in U-Net, which simply concatenate low-level and high-level features without fully utilizing spatial relationships, we have incorporated a 3D Spatial Attention Module within the skip connections. This addition enhances the network's ability to focus on crucial spatial relationships in the images, allowing it to effectively capture relevant context and spatial dependencies between features. Through conducting ablation experiments and comparing with state-of-the-art methods, we validate the effectiveness of our design and the superiority of the 3D SA-UNet.\\

\section{Methods}
\subsection{3D Spatial Attention Module}
\begin{figure}
\includegraphics[width=1\textwidth]{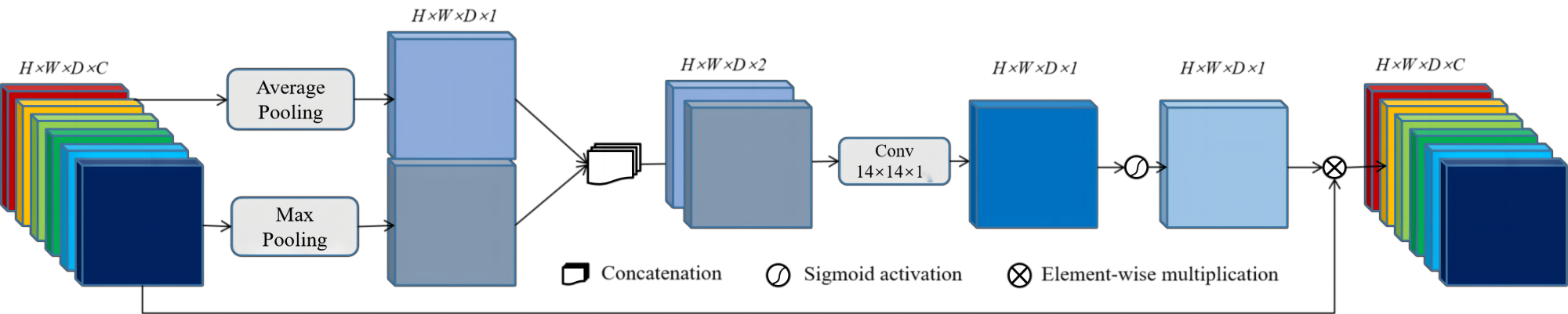}
\caption{Diagram of the 3D Spatial Attention Module.} 
\label{fig1}
\end{figure}

Spatial Attention mechanisms enable the network to amplify significant features and dampen or diminish insignificant ones. Spatial Attention Module is proposed as part of convolutional block attention module (CBAM)\cite{cbam}, and SA-UNet \cite{SAUNET} applies spatial attention to  retinal vessel segmentation in 2D fundus image, making the segmentation performance reached state-of-the-art. Inspired by the above work, we introduce a 3D Spatial Attention Module (3D SAM) to improve the WMH segmentation performance of 3D MRI images. In order to obtain the spatial attention map, 3D SAM utilizes both max-pooling and average-pooling operations along the channel axis. These operations are then concatenated to generate an optimized feature descriptor, as depicted in Fig \ref{fig1}. Mathematically, the input feature map $F\in {R^{H \times W \times D \times C}}$ undergoes channel-wise average-pooling and max-pooling operations, resulting in the generation of outputs${F_{ap}} \in {R^{H \times W \times D \times 1}}$  and ${F_{mp}} \in {R^{H \times W \times D \times 1}}$  , respectively. Afterwards, a convolutional layer is applied to the concatenated feature descriptor, and the output is then passed through the Sigmoid activation function to generate a 3D spatial attention map ${M^{sa}} \in {R^{H \times W \times D \times 1}}$. In short, the output feature map ${F^{sa}} \in {R^{H \times W \times D \times C}}$  of 3D SAM can be calculated as:

\begin{equation}
\begin{split}
F^{sa} & = F \times M^{sa} \\
       & = F \times \sigma({f^{14 \times 14 \times 1}}([F_{ap};F_{mp}])) \\
       & = F \times \sigma({f^{14 \times 14 \times 1}}([Avg(F);Max(F)]))
\end{split}
\end{equation}
where ${f^{14 \times 14 \times 1}(\bullet)}$ refers to a 3D convolutional layer, and $\sigma ({\bullet}) $ means Sigmoid function.

\subsection{3D Atrous Spatial Pyramid Pooling}
Atrous Spatial Pyramid Pooling (ASPP) is a widely utilized module in deep learning models, initially introduced in the DeepLab \cite{deeplab} series of studies for the task of semantic segmentation in 2D images. The core strength of ASPP lies in its ability to process features at different scales. By applying convolutional kernels with varying dilation rates to the output feature maps of a Convolutional Neural Network (CNN), ASPP achieves an aggregation of multi-scale information. This fusion of multi-scale features significantly enhances the model's ability to recognize and segment objects of varying sizes within images.

\begin{figure}
\includegraphics[width=1\textwidth]{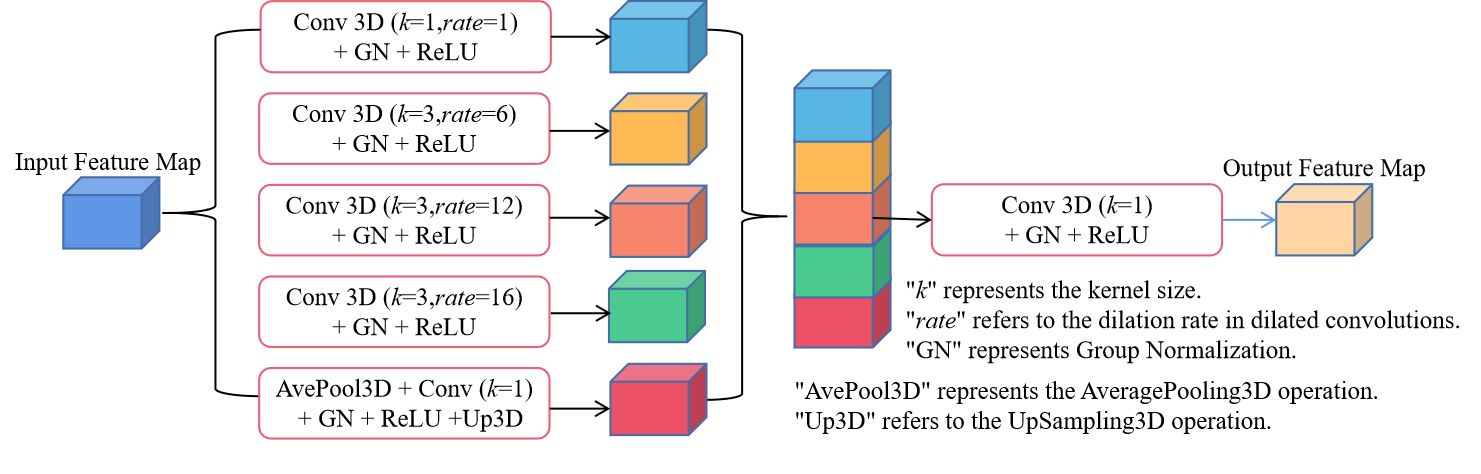}
\caption{Diagram of the 3D Atrous Spatial Pyramid Pooling (3D ASPP).} 
\label{fig2}
\end{figure}

In subsequent research, to meet the demands of 3D medical image segmentation, the ASPP module has been extended into three-dimensional space, resulting in the 3D ASPP module. This enhancement retains the original ASPP's advantage in fusing multi-scale features while introducing Group Normalization (GN) \cite{gn} to replace traditional Batch Normalization (BN), further improving the model's adaptability and stability for small batch sizes. The 3D ASPP captures multi-scale contextual information from local to global through parallel convolutional operations, with each convolutional branch applying a different dilation rate, as shown in Fig \ref{fig2}.

We believe that the introduction of 3D ASPP can significantly enhance the model's understanding of 3D spatial structures, especially when processing complex medical images, enabling more accurate identification and segmentation of key anatomical structures and lesion areas. This in-depth comprehension of spatial context gives 3D ASPP a significant advantage in improving segmentation accuracy, bringing new technological breakthroughs and application potential to the field of medical image analysis.

\subsection{3D SA-UNet}
\begin{figure} 
    \centering
    \includegraphics[width=1\textwidth]{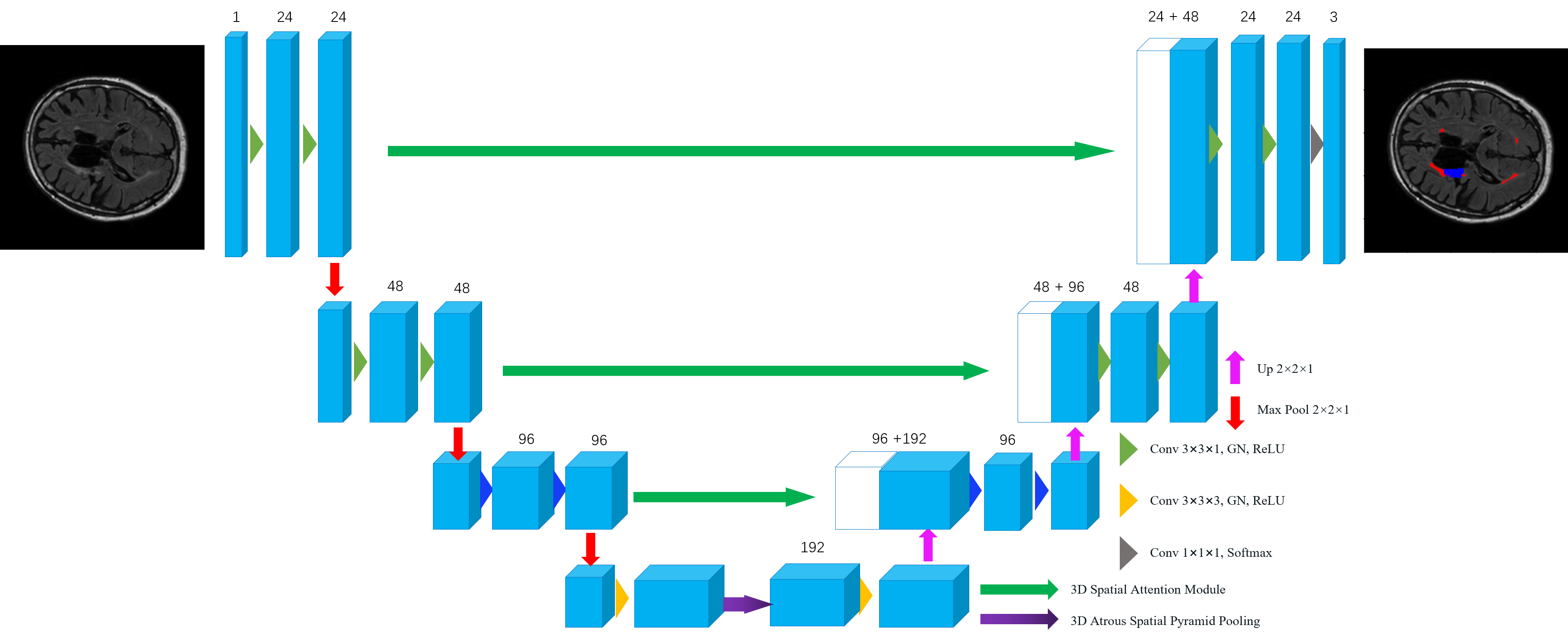} 
    \caption{Diagram of the 3D SA-UNet.} \label{fig3}
\end{figure}

As with the 3D U-Net architecture, our model also adopts the classic encoder-decoder structure, as illustrated in Fig \ref{fig3}. To efficiently handle a minibatch size of 4 for this task, we incorporate Group Normalization (GN) between the convolutional layer and the ReLU layer, as GN has shown to exhibit smaller errors with this minibatch size. Considering the poor imaging resolution along the $z$-axis and the large spatial resolution variations in WMH images, the convolution layers in both the encoder and decoder parts of the model use a kernel size of 3 $\times$ 3 $\times$ 1. However, the bottleneck block between the encoder and decoder retains the use of a 3 $\times$ 3 $\times$ 3 convolution kernel to preserve the network's ability to capture essential 3D features. Similarly, both the encoder's down-sampling layer and the decoder's up-sampling layer are implemented using a kernel size of 2 $\times$ 2 $\times$ 1. Additionally, we introduce a 3D ASPP between the encoder and decoder, which enables the network to learn and integrate multi-scale information more effectively. Moreover, we extend the utilization of 3D SAM to the skip connections, aiming to enhance the network's ability to focus on crucial spatial relationships in the images. This allows the model to effectively capture relevant context and spatial dependencies between features.

\section{Experiments and Results}
\label{sec:pagestyle}
\subsection{Datasets}

To evaluate the performance of our proposed 3D SA-UNet, we conducted experiments on the MICCAI 2017 WMH Segmentation Challenge Dataset, specifically for the task of White Matter Hyperintensities (WMH) segmentation. This challenge was jointly initiated by the UMC Utrecht, NUHS Singapore, and VU Amsterdam, with the aim of benchmarking automatic WMH segmentation methods presumed to be of vascular origin.We exclusively used the FLAIR images from this dataset for our experiments, with 60 images from three scanners for training and 110 images from five different scanners for testing, as briefly summarized in Table \ref{tab:tab1}. Among the test scanners, three of them correspond to the same scanners utilized during training, while the remaining two scanners were not employed in the training phase. Each image is accompanied by manually annotated ground truth, with labels divided into 0 for background, 1 for WMH, and 2 for other pathologies, for further details about the dataset, please refer to MICCAI 2017 WMH Segmentation Challenge\cite{Kuijf}. 

\subsection{Experimental Setup}
In Table \ref{tab:tab1}, it is observed that the maximum dimensions of the training dataset are 240 in height, 256 in width, and 83 in depth. To preserve all the original information, we have applied zero-padding around all sides, adjusting the dimensions to 256$\times$256$\times$128. For the test set, during testing with the specific scanner ``3T Philips Ingenuity" of VU Amsterdam, we used padding/cropping to achieve a size of 256$\times$256$\times$128. We thoroughly examined all the lesion regions within this area of 256$\times$240$\times$83. Therefore, restoring to the dimensions of 321$\times$240$\times$83 is reasonable. Before inputting into the network for training and inference, each 3D image is sliced along the $z$-axis into 4 parts, meaning the input image size to the network is 256$\times$256$\times$32. During testing, the images are concatenated back to the size of 256$\times$256$\times$128 and ultimately restored to their original dimensions.

To enhance the robustness of the network, we employ multiple augmentation techniques, including rotation, random flipping, and random transposition, as well as channel shift, bias field correction, elastic deformations, and motion ghosting. All models in this paper were trained from scratch using a combined loss function (Cross-entropy and Dice Loss) and optimized with the Adam method, initialized with a learning rate of 0.001. 
\subsection{Evaluation Metrics}
In this paper, we have selected three distinct metrics to assess the segmentation performance from various perspectives. Given the ground-truth G and the predicted segmentation map P generated by our model, the three evaluation metrics are defined as follows:
\subsubsection{Dice Similarity Coefficient (DICE)}
\begin{align}
    DICE=\frac{2(P\cap G)}{\left | P \right | + \left | G \right | }   \notag \tag{2}
\end{align}
This quantifies the degree of concordance between P and G in terms of percentage overlap.
\subsubsection{Aerage Volume Difference (AVD)}

Let the lesion region volumes in $P$ and $G$ be denoted as $V_{P}$ and $V_{G}$ , respectively. Consequently, the Average Volume Difference (AVD), is formulated as follows:
\begin{align}
AVD= \frac{\left | V_{P} -V_{G}  \right | }{V_{G}}  \notag \tag{3}
\end{align}
\subsubsection{F1-score for Individual Lesions} 
Let the count of accurately identified lesions in $P$ when contrasted with $G$ be represented by $N_{T}$. Let $N_{F}$ represent the count of erroneously detected lesions within $P$. With each lesion being recognized as a three-dimensional contiguous structure, the F1-score for the discrete lesions is then articulated as:
\begin{align}
F1=\frac{N_{T}}{N_{T}+N_{F}}   \notag \tag{4}
\end{align}
The detailed source code for the computation of the assessment metrics is available in $link^{2}$.
\begin{table}[htbp]  
  \centering  
  \caption{Brief Summary of the MICCAI WMH Challenge dataset}  
  \begin{tabular}{ccccc}  
    \toprule  
    \textbf{Institute} & \textbf{Scanner}& \textbf{Size} & \textbf{Train} & \textbf{Test} \\  
    \midrule  
    UMC Utrecht & 3T Philips Achieva & 240 $\times$ 240  $\times$ 48  &20 & 30  \\  
    NUHS Singapore & 3T Siemens TrioTim &232 $\times$ 256  $\times$ 48  &20 & 30   \\  
    \multirow{3}{*}{VU Amsterdam} & 3T GE Signa HDxt &132 $\times$ 256  $\times$ 83  &20 & 30  \\
    & 3T Philips Ingenuity &321 $\times$ 240  $\times$ 83  & 0 & 10 \\
    & 1.5 T GE Signa HDxt &128 $\times$ 256 $\times$ 103& 0 & 10 \\
    \bottomrule  
  \end{tabular}  

  \label{tab:tab1}  
\end{table}
\subsection{Ablation Studies}
To verify the significance of each component in the 3D SA-UNet for this task, we employ the traditional 3D U-shaped network with Batch Normalization and the initial convolutional layer consisting of 24 channels as our Backbone. Firstly, we compare the ``Backbone with (3,3,3) \& (2,2,2)" and the ``Backbone with (3,3,1) \& (2,2,1)". The main difference between them lies in the convolutional kernel size of the encoder and the decoder. The former uses a kernel size of (3,3,3), while the latter uses (3,3,1). Additionally, there is a difference in the size of the down-sampling layers and the up-sampling layers. The former has a size of (2,2,2), whereas the latter has a size of (2,2,1). As we observe from the results in the first two rows of table \ref{tab:tab2}, ``Backbone with (3,3,1) \& (2,2,1)" compared to ``Backbone with (3,3,3) \& (2,2,2)" was significantly improved in all metrics, especially F1, which improved by more than 20\%, and these findings substantiate that the adjustments align with the characteristics of the input data.  Secondly, we compare the ``Backbone with (3,3,1) \& (2,2,1)" and the ``Backbone* with GN" to validate the effectiveness of Group Normalization (GN). Here, ``Backbone* with GN" represents the basic architectural design that is identical to ``Backbone with (3,3,1) \& (2,2,1)" but with the replacement of BN by GN. From the results, it can be observed that ``Backbone* with GN" achieves better performance in all metrics compared to ``Backbone with (3,3,1) \& (2,2,1)", indicating that Group Normalization (GN) is beneficial for this task. Thirdly, we individually integrate 3D SAM and 3D ASPP into the ``Backbone* with GN", and the outcomes indicate that both 3D ASPP and 3D SAM contribute to enhancing segmentation performance, especially the introduction of 3D SAM, which increased the F1 score from 0.71 to 0.75. Finally, through the combination of 3D SAM and 3D ASPP, our 3D SA-Net is finally built, which achieves the highest DICE of 0.79 and F1 of 0.76, although AVD is not the lowest, it is also comparable. These results demonstrate the effectiveness of the 3D SA-UNet design for this task. We are confident that the evidence presented sufficiently illustrates the effectiveness of the 3D SA-UNet design in this study.

\begin{table}[htbp]  
  \centering  
  \caption{Ablation Experiments on MICCAI Challenge}  
  \begin{tabular}{cccc}  
    \toprule  
     \textbf{Models} & \textbf{DICE $\uparrow$} & \textbf{AVD $\downarrow$} & \textbf{F1$\uparrow$} \\  
    \midrule  
   Backbone with (3,3,1) \& (2,2,1) & 0.75  & 0.225  &   0.68 \\  
    Backbone with (3,3,1) \& (2,2,1) & 0.75  & 0.225  &   0.68  \\  
    Backbone* with GN  & 0.78& 0.172 & 0.71 \\
    Backbone* with GN + 3D ASPP & 0.77  &\textbf{0.170} & 0.71 \\
    Backbone* with GN + 3D SAM & 0.78 & 0.182  & 0.75\\
    3D SA-UNet & \textbf{0.79}  &  0.174 &   \textbf{0.76}\\

    \bottomrule  
  \end{tabular}  
  \label{tab:tab2}  
\end{table}

\begin{figure}
\includegraphics[width=1\textwidth]{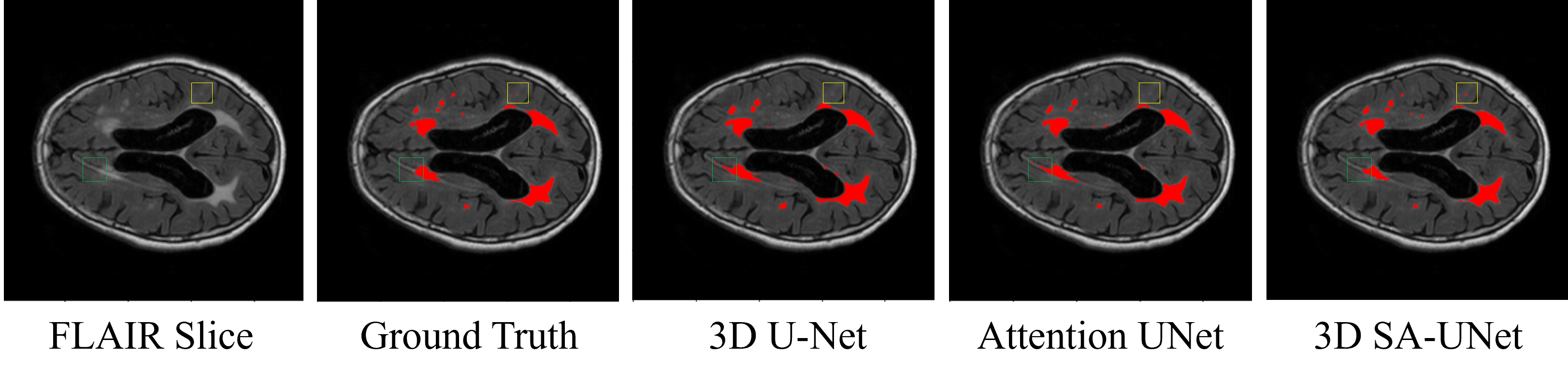}
\caption{Visual comparison of segmentation results for one sample from the testset. 
The area enclosed in the yellow box indicates that the 3D SA-UNet has stronger small-volume WMH recognition capabilities, while the area enclosed in the green box shows that the 3D SA-UNet is less prone to false segmentation compared to the 3D U-Net and the Attention UNet} \label{fig4}
\end{figure}

\subsection{Comparison with other state-of-the-art models}
It is widely acknowledged that since the advent of U-Net, numerous neural networks for medical image segmentation have emerged as adaptations of the U-Net framework. This trend extends to the domain of 3D imaging as well, where a variety of U-Net-inspired 3D models have been introduced, such as 3D-UNet\cite{3dunet}, Attention UNet\cite{attunet}, and others. These models harness the foundational architecture of U-Net, integrating 3D convolutions and attention mechanisms to enhance the accuracy and efficiency of medical image segmentation.

Moreover, we have compared our outcomes with those of the Challenge Winner\cite{Li}, who employed an ensemble of 2D U-Nets with both T1 and FLAIR imaging modalities and achieved the top performance in the White Matter Hyperintensity (WMH) Segmentation Challenge. Quantitatively, as illustrated in Table \ref{tab:tab3}, our proposed method achieves state-of-the-art performance. When juxtaposed with 3D U-UNet and Attention UNet, 3D SA-UNet surpasses them in all evaluation metrics. In contrast to the Challenge Winner, 3D SA-UNet achieves the smallest Average Volume Difference (AVD) of 0.174, a reduction of nearly 4.5\%, matches the F1 score of 0.76, and offers a comparable Dice coefficient of 0.79.

Visually, as depicted in Figure \ref{fig4}, 3D U-Net and Attention UNet struggle with identifying small-volume WMH areas and exhibit missegmentation. In contrast, 3D SA-UNet accurately identifies a greater number of small-volume WMH regions and displays fewer instances of missegmentation. This indicates that 3D SA-UNet possesses higher robustness and precision when dealing with intricate and subtle features in medical imaging.

Furthermore, we have presented the performance of the 3D SA-UNet on images from a diverse range of five distinct scanners within the test set, as detailed in Table \ref{tab:tab4}. Notably, the model's efficacy is maintained on images from scanners not represented in the training dataset, including the "3T Philips Ingenuity" and "1.5T GE Signa HDxt" from VU Amsterdam. These findings underscore the robustness and adaptability of the 3D SA-UNet across various imaging modalities.

Additionally, the design of 3D SA-UNet takes into account the potential differences in image characteristics that may arise from different imaging devices. By incorporating adaptive feature extraction and attention mechanisms into the network, the model is better equipped to adapt to the texture and contrast of images produced by various scanners. This flexibility ensures that 3D SA-UNet not only excels on a single imaging device but is also applicable in a diverse clinical setting, making it a powerful tool for a wide range of medical image analysis tasks.

\begin{table}[htbp]  
  \centering  
  \caption{Comparison with Other State-of-the-art Models ($^\dagger$ indicates results from the original paper)}  
  \begin{tabular}{cccc}  
    \toprule  
     \textbf{Models} & \textbf{DICE$\uparrow$} & \textbf{AVD $\downarrow$} & \textbf{F1$\uparrow$} \\  
    \midrule  
   3D U-Net \cite{3dunet} &  0.71 & 0.289  & 0.53  \\  
   Attention UNet \cite{attunet} & 0.74    & 0.206  & 0.57  \\  
    MICCAI Challenge Winner \cite{Li} $^\dagger$  & \textbf{0.80} &   0.219 & 0.76 \\
     3D SA-UNet  & 0.79 &  \textbf{0.174}  &   \textbf{0.76} \\

    \bottomrule  
  \end{tabular}  
  \label{tab:tab3}  
  
\end{table}
In summary, 3D SA-UNet has demonstrated exceptional performance across a variety of imaging devices and complex medical image features. These results not only substantiate its effectiveness in the task of white matter hyperintensity (WMH) segmentation but also suggest new possibilities and solutions for future medical image segmentation tasks. We believe that the advantages of 3D SA-UNet will enable it to play an increasingly significant role in the field of medical image analysis, providing vital support for clinical diagnosis and treatment.

\begin{table}[htbp]  
  \centering  
  \caption{Segmentation Performance of 3D SA-UNet on Different Scanners}  
  \begin{tabular}{ccccc}  
    \toprule  
    \textbf{Institute} & \textbf{Scanner} & \textbf{DICE$\uparrow$} & \textbf{AVD $\downarrow$} & \textbf{F1$\uparrow$}  \\  
    \midrule  
    UMC Utrecht & 3T Philips Achieva &  0.79 & 0.151 & 0.71  \\  
    NUHS Singapore & 3T Siemens TrioTim & 0.82 & 0.156 & 0.78  \\  
    \multirow{3}{*}{VU Amsterdam} & 3T GE Signa HDxt & 0.79 &  0.201 &   0.79\\
    & 3T Philips Ingenuity & 0.72 &  0.167 & 0.73  \\
    & 1.5T GE Signa HDxt &  0.77 &  0.225  &   0.80 \\
   
    \bottomrule  
  \end{tabular}  

  \label{tab:tab4}  
\end{table}

\section{Conclusions}
In this paper, we present a novel automated 3D convolutional neural network, referred to as 3D SA-UNet, specifically crafted for White Matter Hyperintensity (WMH) segmentation using only FLAIR scans. We assess the impact of each architectural component through ablation experiments, aiming to gauge the effectiveness of each element within the 3D SA-UNet design. To illustrate the superiority of 3D SA-UNet, we carry out comparative experiments with 3D U-Net and Attention UNet. Additionally, we compare our results to the best-performing method in the MICCAI WMH Segmentation Challenge. Our key findings can be summarized as follows: (1) FLAIR imaging alone proves to be sufficient for accurate WMH segmentation; (2) Properly adapting convolutional kernel sizes based on the input image features is effective, particularly when dealing with the substantial spatial resolution variation along the $z$-axis in 3D medical images; (3) Focusing on spatial correlations and incorporating multi-scale information contributes significantly to enhancing model performance; (4) 3D SA-UNet exhibits robust performance across diverse scanners and imaging protocols; (5) Comparing our results with the state-of-the-art 2D methods (Challenge Winner), we demonstrate that our proposed 3D approach is equally effective for WMH segmentation.

\end{document}